# Decision Support System for Renal Transplantation

Abstract ID: 1511

**MD Ehsan Khan, Avishek Choudhury, Dr. Dae Han Won**
**(Binghamton University, USA)**

**Amy L. Friedman**
**(M.D., Chief Medical Officer & Executive Vice President-LiveOnNy)**

## Abstract

The burgeoning need for kidney transplantation mandates immediate attention. Mismatch of deceased donor-recipient kidney leads to post-transplant death. To ensure ideal kidney donor-recipient match and minimize post-transplant deaths, the paper develops a prediction model that identifies factors that determine the probability of success of renal transplantation, that is, if the kidney procured from the deceased donor can be transplanted or discarded. The paper conducts a study enveloping data for 584 imported kidneys collected from 12 transplant centers associated with an organ procurement organization located in New York City, NY. The predicting model yielding best performance measures can be beneficial to the healthcare industry. Transplant centers and organ procurement organizations can take advantage of the prediction model to efficiently predict the outcome of kidney transplantation. Consequently, it will reduce the mortality rate caused by mismatching of donor-recipient kidney transplantation during the surgery.

## 1. Introduction

Kidney transplantation is a lifesaving intervention for the terminal renal disease. The plea for kidney transplantation overshadows the obtainability of organs; however, about 20% of retrieved kidneys are discarded prior transplantation [1]. Demand for a kidney has always been on the peak. "There were 49,233 registrations on the combined UNOS waiting list as of October 31, 1996, an increase of 207% over December 31, 1988. Of these, 69% were awaiting kidney transplantation" [2]. Since 1998, the frequency of renal transplantation has exceeded the count of 420,118. During 2016, 33,606 renal transplants were recorded, reflecting an 8.5% increase over 2015 and a significant increase of 19.8% since 2012. The development in this domain was encouraged by a growth of 9.2% in the number of deceased donors from 2015 to 2016. Approximately 82% (27,628) of the transplants convoluted organs from deceased donors [3]. The burgeoning need for kidney transplantation mandates immediate attention. To ensure the health and minimized post-transplant death, the paper addresses this concern and attempts to identify factors that determine if the kidney procured from the deceased donor can be transplanted or discarded. Renal transplantation commonly known as kidney transplantation is a crucial treatment for patients with the terminal renal disease. There are a plethora of patients waiting for a kidney transplantation. Moreover, living donor kidney transplantation is habitually unreasonable, and thus patients frequently have protracted waiting times until an ideal deceased donor kidney benefits obtainable. The alarming concern of disrupted kidney donation and transplantation can be addressed by minimizing discard of deceased donor kidneys. However, the features responsible for renal discard are yet to be confirmed. The motivation of this paper is to develop decision support system to enhance allograft and patient outcomes by ensuring desirable utilization of procured kidneys. Customary prognosticators of discard embrace deceased donor age, history of diabetes, terminal creatinine, hypertension, donor type and supplementary components of the Kidney Donor Risk Index (KDRI). Reintroduced consideration has been intensive on stratagems to exemplify the donor pool. Because of the donor dearth, amplifying utility of obtainable organs is progressively important and the paper addresses the same. The other concern that needs attention envelops patient mortality post deceased donor renal transplantation. Research states that patients on dialysis waiting for a transplant possess lower life risk that those who get the wrong kidney. moreover, deficient organ readiness is stemming from augmented waiting times for patients pursuing renal transplants from deceased donors. To ensure ideal kidney donor-recipient match and minimize post-transplant deaths, the paper develops a prediction model that identifies factors that determine the probability of success of renal transplantation, that is, if the kidney procured from the deceased donor can be transplanted or discarded.

The following section describes the problem statement and goal of the paper. With the help of literature reviews, the paper supports its prediction model and highlights its application in Transplant centers and Organ procurement organizations.

## 2. Problem Statement and Objective

There is a pressing need for a greater understanding of the factors that influence mismatching kidney transplantation. Incorrectly matching deceased donor-recipient kidneys to the wrong patient can lead to post-transplant death. Due to the rapid growth in requested kidney transplants and the shortened supply. The paper emphasizes the project around minimizing post-transplant deaths due to mismatched transplant. To ensure ideal kidney donor-recipient match and minimize post-transplant deaths, the paper develops a prediction model that identifies factors that determine the probability of success of renal transplantation, that is, if the kidney procured from the deceased donor can be transplanted or discarded.

## 3 Literature Review

Data Mining Classifications Algorithms for Kidney Disease Prediction in 2015, addressed the use of data mining in the field of renal disease prediction. The research primarily focused on determining the best classification algorithm based on the classification accuracy. The paper concluded SVM to be better than the Naive Bayes classifier algorithm. They also calculated the prediction accuracy to be 70.96 and 76.32 using Naïve Bayes and SVM [4]. Predictors of Deceased Donor Kidney Discard in the United States in 2017, performed a stepwise logistic regression and develop a multivariate risk prediction model for kidney graft discard and authenticated the model. Moreover, the paper found no significant baseline disparities between the training (n = 57474) and validation (n = 14368) units. The multivariate model validation reflected acceptable discriminant function in envisaging kidney discard (AUC = 0.84). The article noticed that the predictors of increased discard enclosed age older than 50 years, enactment of a kidney biopsy, cytomegalovirus seropositive status, donation after cardiac death (DCD), presence of hepatitis B and C seropositive status, cigarette addiction, diabetes, hypertension, terminal creatinine greater than 1.5 mg/dL and AB blood type [1]. Donation After Cardiac Death, The University of Wisconsin Experience with Renal Transplantation in 2003 addressed the concern of shortage of organ donors. The primary motive of the paper was to study and compare donation after cardiac death (DCD) and brain death (DBD). The research found no statistical difference in cold ischemic time, the rate of primary non-function, and graft loss in the first-month post-transplantation. The rate of delayed graft function (DGF) was more for DCD donors (27.5% vs. 21.3%; p = 0.016) and expulsion creatinine was higher for DCD donors (1.92 mg/dL vs. 1.71 mg/dL; p = 0.001). Moreover, they failed to find any significant difference in the rate of medical complications. The research concluded that in long-term the outcome of renal transplantation from DCD donors and DBD donors are equivalent. The article exhibits that kidneys taken from DCD donors possess comparable long-term graft survival to those procured from DBD donors, although with higher rates of delayed graft function. "There was a suggestion of improved survival in the recipients of DBD grafts, however, this did not reach clinical significance, particularly when differences in recipient age, as well as HLA-DR mismatching, were taken into account." [5]. A Comprehensive Risk Quantification Score for Deceased Donor Kidneys, The Kidney Donor Risk Index Background during 2009 advised use of a continuous kidney donor risk index (KDRI) for deceased donor kidneys, conjoining donor and transplant features to enumerate graft failure risk. The article used national data from 1995 to 2005, to explore 69,440 first-time, kidney-only, deceased donor adult transplants. Cox regression was the adopted model that determined the risk of death or graft loss, based on donor and transplant features. The anticipated KDRI incorporates 14 donor and transplant features such as donor age, race, history of hypertension, history of diabetes, serum creatinine, the cerebrovascular cause of death, height, weight, donation after cardiac death, hepatitis C virus status, human leukocyte antigen-B and DR mismatch, cold ischemia time. The article claimed that transplants of kidneys in the peak KDRI quintile had 5-year graft survival of 63%, contrasted with 82% and 79% in the two nethermost KDRI quintiles. The paper confirmed that the classified influence of KDRI on graft outcome is an advantageous decision-making tool at the time of the deceased donor kidney transplant [6]. This paper reviews the above articles to support and contrast with the following findings. The following section explains the dataset used and methods applied to conduct the theoretical research.

## 4. Data Analysis

To understand and explore the dataset better, it is important to do the basic statistical analysis before jumping into the application of machine learning algorithms. Table 1 shows the data statistics and importance for each factor which were calculated using logistic regression. The average CIT arrival hours for a transplanted kidney is 18.43 hours while for discarded kidney it is 21.90 hours. So, higher CIT arrival time means there is a higher chance of discarding the kidney [7]. However, our analysis does not consider CIT arrival as an important feature. Similarly, a higher percentage of glomerulosclerosis and KDPI means there is a higher chance of discarding the kidney procured from the deceased donor. The Table below shows the influencing and unimportant features of the data set. Features with P-value less than 0.05 are accepted as important features.

Table 1: Data Statistics

| Features | P value | Confidence Interval at 95% | Odds Ratio |
| --- | --- | --- | --- |
| Age | 0.0040 | (0.009, 0.045) | 1.0271 |
| Gender | 0.0770 | (-0.032, 0.622) | 1.3426 |
| Per_GS | 0.0000 | (-0.102, -0.044) | 0.9297 |
| Per_KDPI | 0.0040 | (-2.928, -0.574) | 0.1735 |
| Cit_Arrival | 0.0091 | (-0.003, 0.036) | 1.0168 |
| Hist_Diabetes | 0.5870 | (-0.455, 0.803) | 1.1903 |
| Hist_HTN | 0.9770 | (-0.423, 0.411) | 0.9937 |

This paper also determines the ranking of the features within the dataset using forest tree and confirms the importance of KDPI, CIT, AGE, and GS. However, the analysis still neglects the importance of diabetes. Figure 1 below shows the feature rankings.

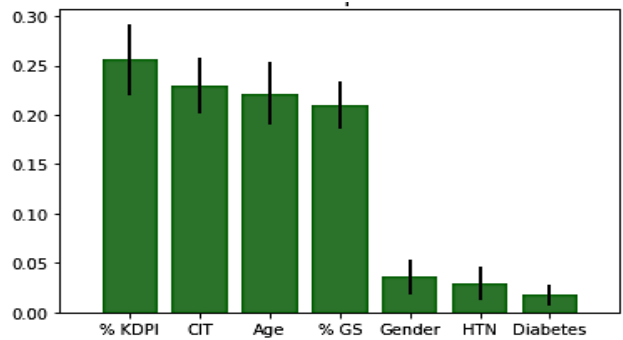

Figure 1: Feature ranking

Variable importance generated by random forest algorithm approximates the importance of a variable by analyzing the prediction error increase when (OOB) data for that feature is permuted leaving the rest unaltered. The required calculations are performed tree by tree as the random forest is made.

## 5. Models and Approaches

### 5.1 Data Preprocessing
Often when performing regression models on a dataset it is likely that it will require some data manipulation. This is done through data imputation, as well as normalizing the data interties. Data imputation is a method for controlling the variance that missing values give a dataset. Datasets are rarely 100% complete, they often will have many missing values. to deal with this data imputation replaces the missing values with the mean of any comparable cases or values. This creates a whole dataset with much less variability. There are many ways to measure the spread of a data set's values. They include variance and standard deviation, however, the method used in the report was the median absolute deviation or MAD [8]. The median absolute deviation is more useful when it is important to reduce the effect of outliers in a dataset. Because MAD is less impacted by median outliers than mean outliers. This works by testing and setting where the cutoff points for the outliers should be set at. Another Approach to correcting this issue is through normalization of a data set. Normalizing a dataset between 0 and 1 will even out the varying range of the data. This makes the training process less sensitive to the scaled range of data. Training datasets use the gradient descent method, the speed at which it can calculate the equations depends on the data features. If the data is scaled, gradient descent can perform the calculations much more efficiently. The equation used in this report to normalize the data can be seen below.

$$Normalized(e_i) = \frac{e_i - E_{min}}{E_{Max} - E_{Min}} \quad (1)$$

$$E_{Min} = the\,minimum\,value\,for\,variable\,E$$
$$E_{Max} = the\,maximum\,value\,for\,variable\,E$$

Note that if $E_{Max}$ is equal to $E_{Min}$ then Normalized ($e_i$) is set to 0.5

### 5.2 Performance Measures

When evaluating supervised training results, it is important to check the performance of the training and testing through the accuracy, sensitivity, specificity, and the AUC values. All these performance measures come from equations based on the contingency matrix. This matrix is based on verifying combinations of true and false cases. The cases are true positive TP, true negative TN, false negative FN, and false positive FP. Accuracy shows the percentage of correctly estimated true positive cases in the dataset. Sensitivity shows the percentage of correctly detected positive cases. Specificity shows the percentage of correctly detected negative cases.

$$Accuracy = \frac{TP+TN}{TP+TN+FP+FN} \quad (2)$$

$$Sensitivity = \frac{TP}{TP+FN} \quad (3)$$

$$Specificity = \frac{TN}{TN+FP} \quad (4)$$

### 5.3 Models

The research project employees supervised learning classification algorithms to classify the outcome of the imported kidney. Four predefined algorithms have been used to classify the kidney outcome and the performance measures are calculated to compare the optimal algorithms to accurately predict the kidney outcome. Table 2 on the following page shows the accuracy of different classification algorithms for the testing dataset. The dataset is randomly shuffled eliminating any bias within the data and divided into training and testing. 90% of the data are used to train the models and 10% of the data are used to predict and test the performance of the fitted models. Papers briefly discuss the best four prediction model algorithm.

#### 5.3.1 Gradient Boosting Classifier

The purpose of boosting algorithms is to "boost" the minor advantage that an assumption formed by a weak learner can accomplish by random guessing, by adopting the weak learning process numerous times on a sequence of determined distributions. Boosting methods, notably Adaptive Boost or Gradient Boost, are simple and powerful algorithms. It involves a set of parameters, whose values seemed to be determined in an ad hoc manner. Lately boosting algorithms have been derived as gradient descent algorithms [9-12].

#### 5.3.2 Random Forest

The random forests algorithm abides by the following steps: (a) Draw n-tree bootstrap models from the original dataset. (b) For each of the bootstrap samples, the algorithm burgeons a unpruned classification or regression tree [13]. (c) Predicts fresh data by amassing the estimates of the n-tree trees.

#### 5.3.4 Naive Bayes Algorithm

Naïve Bayes is a generative-based model that produces features independently. The algorithm cogitates an undetermined target function as P(y=x). To learn, P(y=x) is used in training dataset to gauge P(x=y) and P(y) using the following equation.

$$P(Y = yi|X = xk) = \frac{P(X = xk|Y = yi)P(Y=yi)}{\sum P(X = xk|Y = yi)P(Y=yi)} \quad (5)$$

Moreover, the algorithm can construct predictions for the test dataset by contemplating at likelihoods from distributions. Simultaneously, parameters can be estimated using maximum likelihood or Bayesian estimates [14].

### 5.3.5 Logistic Regression
Logistic regression is managed by learning from the function as P(y=x). Y is a discrete value, and X is a vector including discrete or continuous values. The algorithm estimates parameters from the training dataset.

$$log \frac{p(x)}{1-p(x)} = o + x \quad (6)$$

$$P(x; b, w) = \frac{e^{o+x}}{1+e^{o+x}} \quad (7)$$

$$P(Y = 1|X) = \frac{1}{1+e^{wo+\Sigma_{i=1}^{n} wixi}} \quad (8)$$

$$P(Y = 0|X) = \frac{e^{wo+\Sigma_{i=1}^{n} wixi}}{1+e^{wo+\Sigma_{i=1}^{n} wixi}} \quad (9)$$

Logistic regression algorithm concludes probability and classifies the testing value by using threshold. Post optimizing the equation parameters, it can be employed to predict the output of testing dataset [15].

## 6. Results
The model like most other literature recognized Per_KDPI, CIT arrival time, Donor Age and Per_GS to be the most relevant features that determine the outcome of a renal transplantation whereas, the presence of diabetes had no effect on renal transplantation. However, existing literature rank history of diabetes to be the crucial factor that influences renal transplant. The article recognizes Gradient Boosting Classifier and Random Forest to be the best predicting model yielding as the accuracy of 77.40% and 75.34% respectively avoiding any overfitting. Table 2 below shows the accuracy, sensitivity, specificity and AUC for all prediction model analyzed.

Table 2: Prediction model accuracy

| Model | Accuracy (%) | Sensitivity | Specificity | AUC |
|---|---|---|---|---|
| Gradient Boosting Classifier | 77.40 | 0.7865 | 0.7544 | 0.7705 |
| Random Forest | 75.34 | 0.7865 | 0.7018 | 0.7441 |
| Naive Bayes | 72.60 | 0.8202 | 0.5789 | 0.6996 |
| Logistic Regression | 73.29 | 0.8764 | 0.5088 | 0.6926 |

The paper also observes the ROC curve for each prediction model and shows the same in figure 2 below. From the figure 2 below, it can be observed that AUC for Gradient Boosting Classifier outperforms the rest by a significant value.

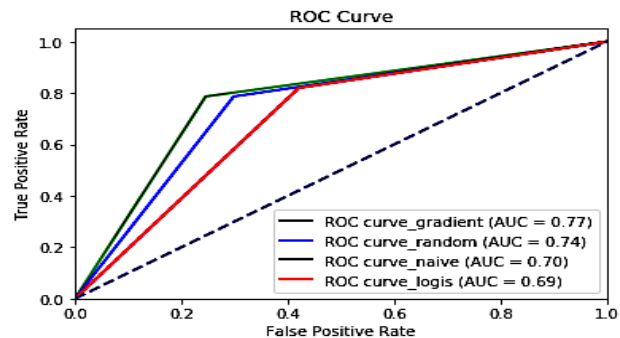

Figure 2: ROC curve and AUC for each model

## 7. Conclusion and Future Scope

The model developed by the paper identifies Donor Age, Per_GS, Per_KDPI and CIT arrival time to be the most crucial factor determining the outcome of a deceased donor renal transplantation. With an accuracy of 77.40% and 75.34%, the paper successfully recommends the use of Gradient Boosting Classifier and Random Forest predicting model over other algorithms for similar data set. The accuracy obtained outperforms the prediction of existing literature in the similar domain which was found to be 76.32 using SVM [5]. The results of this research complement the existing similar literature. The model used in the paper is applicable only to a similar dataset of a deceased kidney donor and might not hold good otherwise. However, the paper contradicts some existing literature and concludes that presence of Diabetes as an insignificant feature for developing decision support system for renal transplantation. Thus, the paper looks forward to further study the effect of diabetes on kidney transplantation. The article requires studying the difference between DBD and DCD donors and its impact on decision support system along with mortality rate of the patient's post-renal transplantation and factors influencing patient health in long-term [16, 17].

## Références

1. Wesley J. Marrero, A. S. (2017). Predictors of Deceased Donor Kidney Discard in the United States. Transplantation, 1690-1697.
2. Harper AM, Rosendale JD, McBride MA, Cherikh WS, Ellison MD: The UNOS OPTN waiting list and donor registry. Clinical Transplant 73–90, 1998.
3. UNOS. (2017, January 15). Retrieved from https://unos.org/data/
4. Dr. S. Vijayarani, M. (2015). Data Mining Classification Algorithms for Kidney Disease Prediction. International Journal on Cybernetics & Informatics, 13-24.
5. Jeffrey T. Coopera, L. T. (2004). Donation After Cardiac Death: The University of Wisconsin Experience with Renal Transplantation. American Journal of Transplantation, 1490-1494.
6. Panduranga S. Rao, D. E. (2009). A Comprehensive Risk Quantification Score for Deceased Donor Kidneys: The Kidney Donor Risk Index. Clinical and Translational Research, 231-236.
7. Kayler, Liise K., Michelle Lubetzky, Xia Yu, and Patricia Friedmann. "Influence of Cold Ischemia Time in Kidney Transplants from Small Pediatric Donors." Transplantation direct 3, no. 7 (2017).
8. Leys, Christophe, Christophe Ley, Olivier Klein, Philippe Bernard, and Laurent Licata. "Detecting outliers: Do not use standard deviation around the mean, use absolute deviation around the median." Journal of Experimental Social Psychology 49, no. 4 (2013): 764-766.
9. Breiman, L. (1997). Prediction games and arcing classifiers. TR 504. Statistics Dept., UC Berkeley.
10. Schapire, R. E. & Singer, Y. (1998). Improved boosting algorithms using confidence-rated predictions. In Proceedings of COLT, 11.
11. Friedman, Jerome, Trevor Hastie, and Robert Tibshirani. "Additive logistic regression: a statistical view of boosting (with discussion and a rejoinder by the authors)." The annals of statistics 28.2 (2000): 337-407.
12. Mason, L., Baxter, J., Bartlett, P., & Frean, M. (1999). Boosting algorithms as gradient descent in function space. In NIPS 11.
13. Liaw, A., & Wiener, M. (2002). Classification and regression by randomForest. R News, 2(3), 18-22.
14. Dai, W., Brisimi, T. S., Adams, W. G., Mela, T., Saligrama, V., and Paschalidis, I. C., 2015, "Prediction of HospitalizationDue to Heart Diseases by Supervised Learning Methods, " International Journal of Medical Informatics, 84(3), 189-197.
15. Mitchell, T. M., 2016, "Generative and Discriminative Classifiers: Naive Bayes and Logistic Regression, " retrieved from http://www.cs.cmu.edu/ tom/mlbook/NBayesLogReg.pdf
16. Der Werf WJ, D'Alessandro AM, Hoffmann RM, Knechtle SJ. Procurement, preservation, and transport of cadaver kidneys. Surg Clin North Am 1998; 78: 41–54. 2. Kootstra G, Kievit J, Nederstigt A. Organ donors: heartbeating and non-heart beating. World J Surg 2002; 26: 181–184.
17. Wolfe RA, Ashby VB, Milford EL, Ojo AO, Ettnger RE, Agodoa LYC, Held PJ, Port FK: Comparison of mortality in all patients on dialysis, patients on dialysis awaiting transplantation, and recipients of a first cadaveric transplant. New England J Med 341: 1725–1730, 1999.